\documentclass
[superscriptaddress,secnumarabic,amssymb,amsmath,nobibnotes,aps,prd,showkeys,showpacs,twocolumn,nofootinbib]{revtex4}%
\usepackage{bm}
\usepackage{hyperref}
\usepackage{mathrsfs}
\usepackage{xcolor,color,graphicx,graphics}
\usepackage[all]{xy}
\usepackage{latexsym,amssymb,amsmath,amsfonts} 
\usepackage[english]{babel} 
\usepackage[OT1]{fontenc}
\usepackage[latin1]{inputenc}
\usepackage{makeidx}
\usepackage{hyperref}
\usepackage{color,graphicx,graphics,wrapfig,epsf}
\usepackage{subfig}
\usepackage{mwe}
\usepackage{orcidlink} 
\hypersetup{urlcolor=BlueViolet,
	    citecolor=Plum,
	    linkcolor=PineGreen}

\begin{document}


\title{Unveiling Phase Space Modifications: A Clash of Modified Gravity and the Generalized Uncertainty Principle}


\author{Aneta Wojnar\orcidlink{0000-0002-1545-1483}}
\thanks{Corresponding author}
\email[E-mail: ]{awojnar@ucm.es}
\affiliation{Department of Theoretical Physics \& IPARCOS, Complutense University of Madrid, E-28040, 
Madrid, Spain}

\begin{abstract}
This study explores the link between Modified Gravity and modifications of phase space volume. Analyzing Fermi gas modifications {in the non-relativistic limit of the} Ricci-based gravities, we derive a generalized partition function in the grand-canonical ensemble, connecting Modified Gravity models with the Generalized Uncertainty Principle. Using this correspondence, we also establish bounds on the linear Generalized Uncertainty Principle: $-6\times10^{22}\lesssim\sigma\lesssim 3\times 10^{22}{\text{ s}}/{\text{kg m}}$, as well as lower bounds for Palatini $f(R)$ gravity $\beta > -7.51587\times 10^7 \text{ m}^2$ and Eddington-inspired Born-Infeld gravity $\epsilon > -1.88\times 10^7 \text{ m}^2$, ensuring microscopic stability. This connection also facilitates testing gravity proposals through tabletop experiments.
\end{abstract}

\maketitle

\section{Introduction}

Studying modified gravity is crucial despite the remarkable predictive power of General Relativity (GR), which successfully describes various phenomena, from Solar System dynamics to gravitational waves' detection from black hole \cite{abbott2016observation} and neutron star mergers \cite{abbott2017gw170817}. While GR excels in many areas, it faces limitations in explaining dark matter in astrophysics \cite{rubin1980rotational}, the accelerated cosmic expansion (dark energy) \cite{huterer1999prospects}, and the enigmatic early cosmological inflation \cite{copeland2006dynamics,nojiri2007introduction,nojiri2017modified,nojiri2011unified,capozziello2008extended,CANTATA:2021ktz}. Addressing these challenges necessitates exploring Modified Gravity theories (MG), opening avenues to deepen our understanding of fundamental cosmic phenomena. Furthermore, exploring the gravitational parameter space has recently uncovered untested regions corresponding to galaxies and stellar objects \cite{baker2015linking}. These untapped domains, distinguishing small-scale systems from the cosmological regime, present an opportunity to gain insights into corrections to GR.

The consistency of the equations in MG comes into question due to various indications from previous studies. Notably, chemical potential's dependence on gravity suggests that alterations in the gravitational field description would impact it \cite{kulikov1995low}. Modified Gravity was shown to transform the geodesic deviation equation on a star's surface, resembling Hook's law and introducing corrections to the polytropic equation of state \cite{kim2014physics}. Microscopic quantities, such as opacity, display modifications, implying an effective treatment \cite{sakstein2015testing}. The laws governing thermodynamics, stellar stability, and Fermi gas properties exhibit corrections originating from gravitational proposals as well \cite{Wojnar:2016bzk,Wojnar:2017tmy,sarmah2022stability,wojnar2023fermi}.

Theoretical descriptions of thermonuclear processes in stars' interiors change under modified gravity, impacting energy generation rate computations \cite{sakstein2015hydrogen,Olmo:2019qsj,crisostomi2019vainshtein,rosyadi2019brown,wojnar2021lithium}. The dependence of elementary particle interactions on the local energy-momentum distributions is introduced by some gravity theories \cite{delhom2018observable}. Specific heats, Debye temperature, and crystallization processes in white dwarfs also depend on the gravity model \cite{2022arXiv221204918K}. Chemical reaction rates, influenced by gravity \cite{lecca2021effects}, are expected to change with modifications to this interaction.

Relativistic effects in equations of state, when ignored in Tolman-Oppenheimer-Volkoff equations derived from GR, lead to an underestimation of compact star limiting masses. Equations of state in curved spacetime for degenerate stars explicitly depend on metric components, resulting in changes to chemical potentials and temperatures \cite{hossain2021equation,hossain2021higher,li2022we,chavanis2004statistical}. Thermodynamic quantities and equations of state are further altered when (pseudo-)scalar fields, like axions, are considered \cite{sakstein2022axion}. 

The Generalized Uncertainty Principle models (GUP) give rise to corrections in equations of state and microscopic variables due to the interplay of special relativity and gravity. This proposition introduces a dispersion relation involving energy, mass, and momentum within Heisenberg's Uncertainty Principle, incorporating the constants representing the speed of light ($c$) and gravity ($G$), the latter symbolizing the Newtonian constant \cite{moussa2015effect,rashidi2016generalized,belfaqih2021white,mathew2021existence,hamil2021new,gregoris2022chadrasekhar}.

Simultaneously, the growing importance of integrating the quantum structure of space-time and deforming associated quantum phase spaces, leading to the generalization of the Heisenberg uncertainty principle, is emphasized for the potential measurable effects it offers \cite{Pachol:2023bkv,Pachol:2023tqa,Kozak:2023vlj}. GUP has proven valuable in predicting quantum gravity effects, as evident in various models found in \cite{kempf1995hilbert,maggiore1993generalized,maggiore1994quantum,chang2002exact,chang2002effect,moussa2015effect}. Despite variations in mathematical structures, a shared characteristic among most of these models is the existence of a minimum length scale, anticipated to be approximately the Planck length, $L_P\sim \sqrt{\frac{\hbar G}{c^3}}$, as detailed in \cite{bishop2020modified,bishop2022subtle,segreto2023extended}.

Subsequently, we will illustrate a correspondence between {the non-relativistic limit of} MG and the GUP. The following section will revisit fundamental concepts associated with a class of metric-affine theories known as Ricci-based gravities. Section 3 will unveil our central outcome, showcasing the affirmed correspondence that yields a modified partition function featuring a deformed weighted phase space volume. In Section 4, we will explore the constraints imposed on Ricci-based gravities and GUP models arising from this correspondence. The concluding section is dedicated to summarizing our findings and presenting our overall conclusions.

\section{Ricci-based gravities}
In the subsequent discussion, we delve into a class of metric-affine gravity proposals known as Ricci-based gravity theories (see, for instance, \cite{alfonso2017trivial}). The action associated with this specific class is expressed as follows:

\begin{equation} \label{eq:actionRBG}
\mathcal{S}=\int d^4 x \sqrt{-g} \mathcal{L}_G(g_{\mu\nu},R_{\mu\nu}) +
    \mathcal{S}_m(g_{\mu\nu},\psi_m)  .
    \end{equation}
Here, $g$ represents the determinant of the space-time metric $g_{\mu \nu}$, and $R_{\mu \nu}$ denotes the symmetric Ricci tensor. Notably, the latter is independent of the metric, constructed solely with the affine connection $\Gamma \equiv \Gamma_{\mu\nu}^{\lambda}$. We introduce the object ${M^ \mu}_{\nu} \equiv g^{\mu \alpha}R_{\alpha\nu}$ to formulate the gravitational Lagrangian $\mathcal{L}_G$, constructed as a scalar function using powers of traces of ${M^ \mu}_{\nu}$.

On a different note, the matter action is given by:

\begin{equation}
\mathcal{S}_m=\int d^4 x \sqrt{-g} \mathcal{L}_m(g_{\mu\nu},\psi_m).
\end{equation}
In this proposal, the matter action is minimally coupled to the metric, disregarding the antisymmetric part of the connection (torsion), akin to the treatment of minimally coupled bosonic fields. This simplification extends to fermionic particles, such as degenerate matter, effectively described by a fluid approach, exemplified by the perfect fluid energy-momentum tensor \cite{alfonso2017trivial}. Similarly, by focusing on the symmetric part of the Ricci tensor, potential ghostlike instabilities are avoided \cite{beltran2019ghosts,jimenez2020instabilities}. This approach accommodates various gravity theories, including GR, Palatini $f(R)$ gravity, Eddington-inspired Born-Infeld (EiBI) gravity \cite{vollick2004palatini}, and its numerous extensions \cite{jimenez2018born}.

The gravitational action encompasses theories that, despite possessing intricate field equations, can be reformulated conveniently, as demonstrated in \cite{jimenez2018born}:

\begin{equation} \label{eq:feRBG}
{G^\mu}_{\nu}(q)=\frac{\kappa}{\vert \hat{\Omega} \vert^{1/2}} \left({T^\mu}_{\nu}-\delta^\mu_\nu \left(\mathcal{L}_G + \frac{T}{2} \right) \right) .
\end{equation}
Here, $\vert\hat{\Omega}\vert$ denotes the determinant of the deformation matrix, and $T$ is the trace of the energy-momentum tensor of matter fields. The Einstein tensor ${G^\mu}_{\nu}(q)$ is associated with a tensor $q_{\mu\nu}$, where the connection $\Gamma$ assumes the Levi-Civita connection of $q_{\mu\nu}$:

\begin{equation}
\nabla_{\mu}^{\Gamma}(\sqrt{-q} q^{\alpha \beta})=0.
\end{equation}
For this formalism, the tensor $q_{\mu\nu}$ is related to the space-time metric $g_{\mu\nu}$ through:
\begin{equation}\label{eq:defmat}
q_{\mu\nu}=g_{\mu\alpha}{\Omega^\alpha}_{\nu}.
\end{equation}
The deformation matrix ${\Omega^\alpha}_{\nu}$ is theory-dependent, determined by the gravitational Lagrangian $\mathcal{L}_G$. Importantly, those theories yield second-order field equations, reducing to GR counterparts in vacuum (${T_\mu}^{\nu}=0$). This implies no extra degrees of freedom propagate in these theories beyond the usual two polarizations of the gravitational field.

Of particular interest is the nonrelativistic limit of the field Eqs. \eqref{eq:feRBG}. In Palatini $f(R)$ \cite{Toniato:2019rrd} and EiBI \cite{banados2010eddington,pani2011compact} gravities, the Poisson equation takes the form:

\begin{equation}\label{poisson}
\nabla^2\phi = \frac{\kappa}{2}\Big(\rho+\alpha\nabla^2\rho\Big).
\end{equation}
Here, $\phi$ is the gravitational potential, $\kappa=8\pi G$, and $\alpha$ is a theory parameter. The expressions for $\alpha$ are $\alpha=2\beta$ for Palatini $f(R)$, with $\beta$ accompanying the quadratic term, and $\alpha=\epsilon/2$ for EiBI, where $\epsilon=1/M_{BI}$ and $M_{BI}$ is the Born-Infeld mass. It is noteworthy that the similarity in the Poisson equation between these two gravity proposals is not coincidental; the EiBI gravity in the first-order approximation reduces to Palatini gravity with the quadratic term \cite{pani2012surface}. Furthermore, only the quadratic term $R^2$ influences the non-relativistic equations, as higher curvature scalar terms enter the equations at the sixth order \cite{Toniato:2019rrd}.

Concerning constraints on Ricci-based gravities, the analysis within the $f( R)= R+\beta  R^2$ model indicates that $|\beta|$ typically remains below $2\times 10^{8}\, \text{m}^2$ \cite{Olmo:2005zr} in the weak-field limit. However, due to uncertainties in microphysics, experiments within the Solar System lack precise constraints on these parameters \cite{Toniato:2019rrd}. Notably, tests of gravity in vacuum, like the Shapiro delay technique, offer no constraints for Palatini gravity, reducing to GR with a cosmological constant, as discussed post Eq. \eqref{eq:defmat}.

Conversely, considering microphysical aspects, seismic data from Earth imposes stricter limitations on the parameter, with $|\beta|\lesssim 10^9\, \text{m}^2$ (at a $2\sigma$ level of accuracy) \cite{Kozak:2023axy,Kozak:2023ruu}. On the other hand, examining neutron stars' equations of state and combining the results with observational data sets a limit at $|\beta| \lesssim 10^6$ m$^2$ \cite{Lope-Oter:2023urz}. Similar to general relativity, neither $f( R)$ models nor EiBI models effectively explain the rotation curves of galaxies \cite{Hernandez-Arboleda:2022rim,Hernandez-Arboleda:2023abv}. Consequently, constraints from galaxy catalogs remain elusive to date.

\section{Thermodynamics and Phase Space of Ricci-Based Gravities}

In \cite{Pachol:2023bkv}, it was demonstrated that a broad class of GUP models\footnote{Derived from the Snyder model \cite{snyder1947quantized} of non-commutative spacetime; see its thermodynamics in \cite{Pachol:2023tqa}.} can be indistinguishable from modifications introduced to Einstein's GR when describing a physical system like a star. Both approaches lead to potential observational effects, encompassing mass, radius, surface temperature, and luminosity. Thus, if such effects are observed in stellar astrophysics, it is crucial to note that the influences of noncommutative space-time in Quantum Gravity may resemble those of Modified Gravity. Furthermore, future Quantum Gravity theories are anticipated to converge towards modified Einstein gravity, making modifications in stellar equations consistent with approaches to Quantum Gravity models unsurprising.

In what follows, we will demonstrate that indeed, Ricci-based gravities also lead to a deformed phase space. This provides a framework to study microphysical effects that could be tested in laboratories, similar to GUP proposals (for a review, see \cite{Bosso:2023aht}).

To show that, let us consider a non-relativistic Fermi gas in the low-temperature limit, $T\rightarrow0$. In the case of the non-relativistic limit of MG, its barotropic equation of state has a polytropic form with the polytropic index $\gamma = 5/3$ and does not depend on a theory of gravity\footnote{However, finite temperature corrections do introduce such a dependence, see \cite{chavanis2020statistical,wojnar2023fermi}} \cite{wojnar2023fermi}:

\begin{equation}
P_p= \frac{1}{20}\left(\frac{3}{\pi}\right)^\frac{2}{3}\frac{h^2}{m_e(\mu_e m_u)^\frac{5}{3}} \rho^\frac{5}{3} =: K \rho^\frac{5}{3},
\end{equation}
where $P_p$ is pressure and $\rho$ is energy density. Additionally, the mean molecular weight per electron is given by
\begin{equation}
\mu_e^{-1}=X+\frac{Y}{2}+(1+X+Y)\left<\frac{Z}{A}\right>
\end{equation}
with $\left<{Z}/{A}\right>$ being the average number of electrons per nucleons in metals, $X$ and $Y$ the mass fractions of hydrogen and helium, respectively, while the other symbols have their standard meaning.

Applying the polytropic EoS to the non-relativistic limit of Palatini $f(R)$ gravity in the spherical-symmetric case\footnote{To consider the EiBI case, one needs to rescale the parameter $\beta$, as explained after the Eq. \eqref{poisson}.}

\begin{equation}\label{pois2}
\nabla^2\phi = \frac{\kappa}{2}\Big(\rho+2\beta\nabla^2\rho\Big)
\end{equation}
and using the hydrostatic equilibrium equation

\begin{equation}\label{hydro}
\frac{d\Phi}{dr}=-\rho^{-1}\frac{dP_p}{dr},
\end{equation}
we can write
\begin{equation}
\frac{1}{r^2}\frac{d}{dr} \left(-r^2\rho^{-1} \left[ \frac{dP_p}{dr} + 8\pi G\beta \rho \frac{d\rho}{dr}
\right]
\right) = 4\pi G \rho.
\end{equation}
Note that above equation can be rewritten as
\begin{equation}
\frac{1}{r^2}\frac{d}{dr} \left(-r^2\rho^{-1} \frac{dP }{dr}
\right) = 4\pi G \rho,
\end{equation}
where the effective pressure is given by
\begin{equation}\label{eosmod}
P = K \rho^\frac{5}{3} + 4\pi G \beta \rho^2.
\end{equation}
Let us write it in a more convenient form for the further analysis:
\begin{equation}
P = K \rho^\frac{5}{3} + \sigma K_2 \rho^2,
\end{equation}
where 
\begin{equation}
\sigma = \frac{4\pi G}{K_2}\beta\,\,\,\,\text{and}\;\;\; K_2 = \frac{3}{\pi} \frac{h^3N_A^2}{m_e \mu_e^2}.
\end{equation}
Using a definition of the electron degeneracy parameter $\Psi$ given as
\begin{equation}\label{degeneracy}
\psi:=\frac{k_{B} T}{E_{F}}=\frac{2 m_{e} k_{B} T}{\left(3 \pi^{2} \hbar^{3}\right)^{2 / 3}}\left[\frac{\mu_{e}}{\rho N_{A}}\right]^{2 / 3},
\end{equation}
where $k_B$ is the Boltzmann constant, $T$ temperature, and $E_F$ Fermi energy,
we can rewrite the pressure \eqref{eosmod} as
\begin{equation}\label{pres}
P =\frac{8\pi}{(2\pi^2\hbar^3)^3}\left[
\frac{2}{5}\frac{(2m_e E_F)^\frac{5}{2}}{6m_e}
+\frac{\sigma}{3} \frac{(2m_e E_F)^3}{8m_e}
\right].
\end{equation}

Since we are dealing with the non-relativistic Fermi gas with energy $E\approx p^2/2m_e$ at $T\rightarrow0$, the Fermi-Dirac distribution function 

\begin{equation}\label{distribution}
f(E)=\left(1+z^{-1} e^{E/k_BT}\right)^{-1},
\end{equation}
where  $z=e^{\mu/k_BT}$ while $\mu$ is the chemical potential,
took the step function form, that is,

\[ f(E)=  \left\{
\begin{array}{ll}
      1 & \mbox{if } E\leq E_F \\
      0 & \mbox{otherwise.} \\
\end{array} 
\right. \]

The derivation of the pressure \eqref{pres} can be traced back to a more general Fermi expression, given by
\begin{equation}
P = \frac{4\pi g_s}{(2\pi^2\hbar^3)^3} \int f(E) \left( \frac{(2m_e E)^\frac{3}{2}}{3} +\sigma \frac{(2m_e E)^2}{4} \right) \frac{dE}{dp} dp,
\end{equation}
where $f(E)$ follows the general form \eqref{distribution}, $g_s$ represents the spin of a particle (for electrons, $g_s=2$), $\frac{dE}{dp}=\frac{c^2p}{E}$, and $E=(p^2c^2+m^2c^4)^{1/2}$. The expression can be further reformulated in terms of momentum $p$ as
\begin{equation}\label{pres2}
P= \frac{ g_s}{(2\pi^2\hbar^3)^3} \int \frac{c^2p}{E}
f(E) \left( \frac{p^{3}}{3} +\sigma \frac{p^4}{4} \right) 4\pi dp,
\end{equation}
where the bracket represents a series expansion at $\sigma=0$ up to linear terms in $\sigma$, as derived from
\begin{equation}
-\left( \frac{\sigma p (\sigma p +2)+ 2\text{ln}(1-\sigma p)}{2\sigma^3} \right) \approx
\left( \frac{p^{3}}{3} +\sigma \frac{p^4}{4} \right) + O(\sigma^2).
\end{equation}
Verification shows that \eqref{pres2} can be derived from the following integral for $b=1$, employing integration by parts:
\begin{equation}\label{pres3}
P= k_B T \frac{ g_s}{a(2\pi^2\hbar^3)^3} \int \text{ln}(1+az e^{-\frac{E}{k_B T}})
\frac{4\pi p^2dp}{(1-\sigma p)^b} ,
\end{equation}
where $a=1$ ($a=-1$) for fermions (bosons). Consequently, as pressure is given by (see e.g., \cite{huang2009introduction})\footnote{Other thermodynamic variables such as the number of particles $n$ and internal energy $U$ can also be obtained from \eqref{partition}:
\begin{equation*}\label{therm}
n=k_B T\frac{\partial}{\partial \mu} \mathrm{ln}Z\mid_{T,V},\;\;\;U=k_B T^2\frac{\partial}{\partial T} \mathrm{ln}Z\mid_{z,V} \ .
\end{equation*}
}
\begin{equation}
P= k_B T \frac{\partial}{\partial V} \text{ln}Z,
\end{equation}
where $\text{ln}Z$ is the partition function in the grand-canonical ensemble, the partition function in three dimensions is expressed as
\begin{equation}\label{partition}
\mathrm{ln}Z = \frac{V}{(2\pi \hbar)^3}\frac{g}{a}\int \mathrm{ln}\left[1+az e^{-E/k_BT}\right] \frac{d^3p}{(1-\sigma p)^{b}} \ ,
\end{equation}
with $V:=\int d^3x$ representing the volume of the cell in configuration space. This function describes a system of $N$ particles with energy states $E_i$
\begin{equation}\label{part}
\mathrm{ln}Z = \sum_i \mathrm{ln}\left[1+az e^{-E_i/k_BT}\right]
\end{equation}
considered in a large volume such that
\begin{equation}
\sum_i\rightarrow \frac{1}{(2\pi \hbar)^3} \int \frac{d^3xd^3p}{(1-\sigma p)^{b}} .
\end{equation}
Similar to the GUP with linear $p-$modifications \cite{cortes2020deformed,ali2009discreteness,ali2011minimal,abac2021modified,vagenas2019gup,tawfik2014generalized}, our approach involves a deformed phase space measure with the deformation parameter $\sigma$, which in the case of GUP it is derived from the use of the Liouville theorem \cite{vagenas2019linear}. This implies that the effective $\hbar$ depends on the momentum $p$ in the generalized uncertainty relation, leading to a $p$-dependent size of the unit cell for each quantum state in phase space. Further exploration along these lines is reserved for future work.

\section{Resulting constraints on GUP and Ricci-based gravity}

Let us briefly examine potential constraints derived from the MG and GUP correspondence. It is important to note that alterations to the pressure resulting from Ricci-based gravities are not arbitrary. Consequently, we can establish a limit on the parameter $\beta$, beyond which microphysical laws would be violated.

Conversely, recent constraints on Ricci-based gravities have emerged through the analysis of seismic data obtained from the Earth \cite{Kozak:2021ghd}. This analysis has led to the imposition of a boundary on the theory parameter, expressed as $-2 \times 10^9 \lesssim \beta \lesssim 10^9  \text{m}^2$ \cite{Kozak:2023axy,Kozak:2023ruu}. Consequently, this constraint can be effectively employed to restrict the linear GUP correction as well.

\subsection{ Ricci-based gravities}
Given that the pressure \eqref{pres} must remain positive to ensure microscopic stability in accordance with Le Chatelier's principle, it follows (bearing in mind that $\sigma = \frac{4\pi G}{K_2}\beta$):
\begin{equation}
\beta > - \frac{8}{5 \pi^2 G} \frac{h^3N_A^2}{m_e \mu_e^2}{(2m_e E_F)^{-\frac{1}{2}}} .
\end{equation}
For a typical white dwarf star with a Fermi energy value of $E_F\sim3$ MeV and $\mu_e=2$, the parameter exhibits a lower bound:
\begin{equation}
\beta > -7.51587\times 10^7  \text{m}^2.
\end{equation}
The physical interpretation is as follows. Eq. \eqref{pres} can also be viewed as a standard polytropic EoS $p=\tilde K\rho^{5/3}$, where $\tilde K$ incorporates the term related to electron degeneracy (or Fermi energy). This can be related to the bulk modulus (incompressibility):
\begin{equation}
B= \frac{dP}{d \mathrm{ln}\rho},
\end{equation}
which describes properties of an isotropic material, such as crystallized cores of white dwarf stars or terrestrial planets. Moreover, it can also be expressed in terms of shear modulus and elastic constants, which appear in Hooke's law \cite{poirier2000introduction}. For our EoS, the bulk modulus is then
\begin{equation}\label{bulk}
B=\frac{5}{3}\tilde K\rho^{5/3},
\end{equation}
and it is modified due to alterations in $\tilde K$ provided by the theory parameters. For incompressible solids, $B=\infty$, while for infinitely compressible ones, $B=0$. Therefore, a positive $\beta$ implies an infinitely compressible solid, while a negative $\beta$ corresponds to the incompressible counterpart.

\subsection{GUP with the linear momentum correction}

Taking into consideration that Ricci-based gravities were constrained within a $2\sigma$ accuracy, where the deformation parameter $\sigma$ is linked to the parameter $\beta$ of MG through
\begin{equation*}
\sigma = \frac{4\pi G}{K_2}\beta,
\end{equation*}
the resulting constraint on $\sigma$ is
\begin{equation}
-6\times10^{22}\lesssim\sigma\lesssim 3\times 10^{22}\frac{\text{s}}{\text{kg m}}.
\end{equation}
In the context of existing literature, \cite{farag2012generalized} proposed an upper bound of $\sigma < 10^{24}$ based on considerations of Landau energy shifts for a particle with mass $m$ and charge $e$ in a constant magnetic field and cyclotron frequency. However, leveraging insights from the gravitational wave event GW150914, \cite{feng2017constraining} revised this upper bound to $\sigma< 1.8\times 10^{20}$.

In another study, \cite{ali2011minimal} established an upper bound of $\sigma< 10^{17}$ that aligns with the predictions of the electroweak theory. Additionally, \cite{tawfik2015review} analyzed the effect of linear GUP on the Lamb shift for a hydrogen atom, resulting in a more stringent upper bound of $\sigma < 10^{12}$.

\section{Conclusions}

Motivated by numerous indications suggesting that Modified Gravity may provide insights into the microscopic properties of matter, as briefly discussed in the introduction, this work establishes a connection between corrections to Einstein's General Relativity and modifications of the weighted phase space volume. The deformation of phase space is well-known to occur due to generalizations of the Heisenberg uncertainty principle, a phenomenon predicted by many Quantum Gravity proposals.

In this study, we scrutinize the modifications to the Fermi gas introduced by {the non-relativistic limit of the} Ricci-based gravities. By interpreting the additional term in the Poisson equation (i.e., quadratic corrections to the gravitational Lagrangian) as a modification of the polytropic equation of state, we successfully derive a more general Fermi equation of state. Consequently, a more comprehensive statistics is obtained, enabling the study not only of fermions but also of bosonic particles. Interestingly, the partition function's form is altered due to the deformed phase space, establishing a connection between Modified Gravity models and Generalized Uncertainty Principle.

This correspondence allows us to leverage methods developed by the Modified Gravity community for constraining and testing effective models of Quantum Gravity. In this context, we utilize constraints on Ricci-based gravities derived from earthquake data, establishing a bound on the linear Generalized Uncertainty Principle approach as $-6\times10^{22}\lesssim\sigma\lesssim 3\times 10^{22}{\text{ s}}/{\text{kg m}}$.

Moreover, by understanding the gravitational effects on matter properties, we impose constraints on the parameter $\beta$ of Ricci-based gravities, revealing that $\beta > -7.51587\times 10^7 \text{ m}^2$ for Palatini $f(R)$ gravity ($\epsilon > -1.88\times 10^7 \text{ m}^2$ for Eddington-inspired Born-Infeld gravity) to ensure microscopic stability of matter. This sheds light on the existence of singular values of the theory parameter \cite{kozak2021invariant}, which are dependent on a given equation of state \cite{Lope-Oter:2023urz}.

Let us emphasize that the thermodynamics derived from this correspondence should be applied in the case of non-relativistic objects, such as planets, brown dwarfs, active stars, and white dwarfs. The latter can still be considered in a non-relativistic regime due to their size. Therefore, the compactness for a non-rotating astrophysical object, applicable to our formalism, is $\mathcal{C} << 1$. On the other hand, general Fermi equation of state with modifications introduced by modified gravity in full relativistic regime was provided in \cite{wojnar2023fermi}.

This connection between Modified Gravity and Generalized Uncertainty Principle models opens avenues for testing gravity proposals through various tabletop experiments. Ongoing research along these lines aims to further explore and validate the implications of Modified Gravity on the microscopic properties of matter.

\section*{Acknowledgements}
 AW acknowledges financial support from MICINN (Spain) {\it Ayuda Juan de la Cierva - incorporaci\'on} 2020 No. IJC2020-044751-I. The author expresses gratitude to Anna Pacho\l\, for inspiring discussions.

\bibliographystyle{apsrev4-1}
\bibliography{biblio}

\end{document}